# Captured Small Solar System Bodies in the Ice Giant Region

Community Science White Paper for the Planetary and Astrobiology Decadal Survey, 2023-2032

Timothy Holt[1,2], (Center for Astrophysics, University of Southern Queensland, 720-483-9515, timothy.holt@usq.edu.au )

Bonnie Buratti[7], Julie Castillo-Rogez[6], Björn J. R. Davidsson[7], Tilmann Denk[4], Jonti Horner[1], Bryan J. Holler[10], Devanshu Jha[9], Alice Lucchetti[8], David Nesvorny[2], Maurizio Pajola[8], Simon Porter[2], Alyssa Rhoden[2], Steven Rappolee, Rebecca Schindhelm[5], Linda Spilker[7], Anne Verbiscer[3]

[1] Center for Astrophysics, University of Southern Queensland (CFA-USQ), Towoomba, QLD, Australia;
[2] Department of Space Studies, Southwest Research Institute, Boulder, CO, USA
[3] Department of Astronomy, University of Virginia, Charlottesville, VA, USA.
[4] DLR (German Aerospace Center), Berlin, Germany.
[5] Ball Aerospace, Boulder, CO. USA.
[6] Florida Space Institute, University of Central Florida, USA
[7] NASA's Jet Propulsion Laboratory (JPL), California Institute of Technology, USA
[8] INAF - Astronomical Observatory of Padova, Padova, Italy
[9] MVJ College of Engineering, India
[10] Space Telescope Science Institute, Baltimore, MD, USA

**Executive Summary:** This white paper advocates for the inclusion of small, captured Outer Solar system objects, found in the Ice Giant region in the next Decadal Survey. These objects include the Trojans and Irregular satellite populations of Uranus and Neptune. The captured small bodies provide vital clues as to the formation of our Solar system. They have unique dynamical situations, which any model of Solar system formation needs to explain. The major issue is that so few of these objects have been discovered, with very little information known about them. The purpose of this document is to prioritize further discovery and characterization of these objects. This will require the use of NASA and NSF facilities over the 2023-2032 decade, including additional support for analysis. This is in preparation for potential future *in-situ* missions in the following decades.

## Scope

This whitepaper advocates for the inclusion of small, captured Outer Solar system objects, found in the Ice Giant region in the 2023-2032 The Planetary Science and Astrobiology Decadal Survey. These objects include the Trojans and irregular satellite populations of Uranus and Neptune. This white paper is the companion to others submitted by the Outer Planet Assessment Group (OPAG), particularly those on other small outer Solar System bodies (Omurhan et al, 2020), Uranus (Cartwright et al., 2020) and Neptune (Hofstadter et al., 2020) exploration. This whitepaper is also inter-community, linking to the Small-Body Assessment Group (SBAG) Goals (Swindle et al. 2020), specifically, Goal 1: Small Bodies, Big science, with Objectives 1.1 and 1.2 important, explored further in a related whitepaper (Davidsson et al. 2020) .

## Motivation and Context

This document advocates for the further study of captured small Solar system bodies, Trojans and Irregular satellites, in the context of Uranus and Neptune, the Ice Giants of our Solar system. The other small satellites are covered in a related whiptepaer (Buratti et al. 2020). These objects are within the purview of The Planetary Science and Astrobiology Decadal Survey 2023-2032, specifically Scope Section 1. These objects are of major scientific interest, and the recommendation is for discovery and characterization, in preparation for future missions, as per Considerations in the Decadal Statement of Task, NASA recommendations, section 2 and 3. The NSF-funded Legacy Survey of Space and Time (LSST) at the Vera C. Rubin Observatory will also form a major discoverer of these objects in the next decade, providing opportunities for inter-departmental collaborations.

Additionally, it is critical that the planetary science community fosters an interdisciplinary, diverse, equitable, inclusive, and accessible environment. We strongly encourage the decadal survey to consider the state of the profession and the issues of equity, diversity, inclusion, and accessibility—not as separable issues, but as critical steps on the pathway to understanding captured Small Solar system bodies in the ice Giant Region and the entire solar system. Background information on the current lack of diversity in our community and specific, actionable, and practical recommendations can be found in (Rivera-Valentín et al. 2020, Rathbun et al., 2020, Strauss et al. 2020, Milazzo et al. 2020).

**Trojans of Uranus and Neptune**

Collectively, Trojans are swarms of small Solar system bodies that have been captured at the L4 and L5 Lagrange points of the Giant planets. The most numerous and well studied of these are the Jovian Trojans, with over 7000 objects discovered. The current canonical history of these objects is that they were captured during a period of instability, early in Solar system history

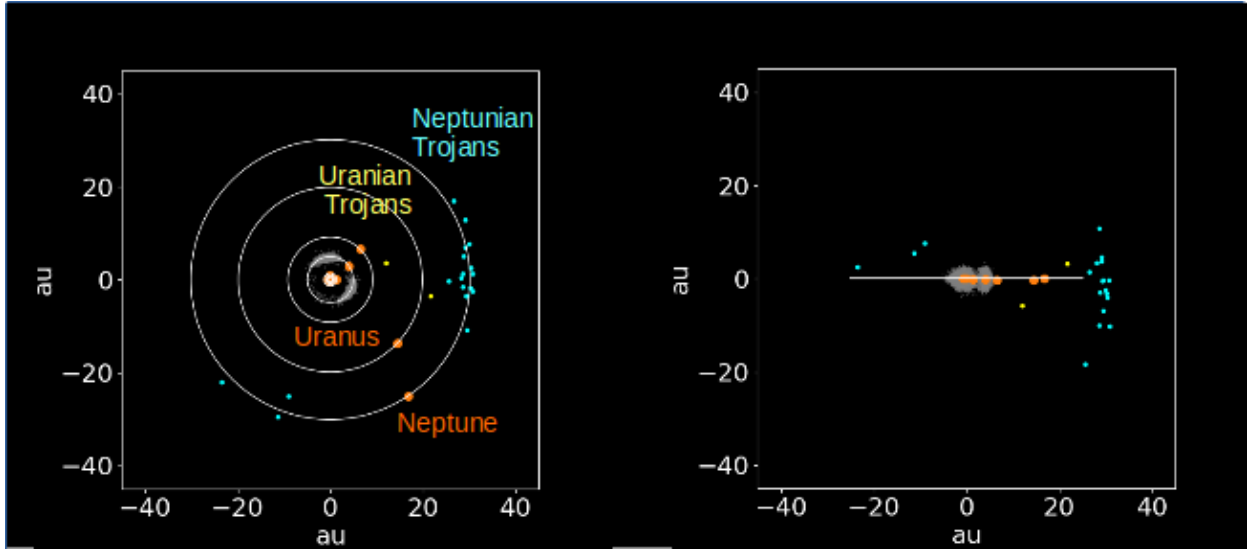

(Lykawka, & Horner, 2010, Nesvorny et al. 2013). During this same instability, and interactions with Jupiter, Saturnian Trojans are dynamically unstable (Nesvorny & Dones, 2002). As with the Saturnian Trojans, the primordial Uranian Trojans deplete quickly (Alexandersen et al., 2014, de la Fuente & de la Fuente, 2014). The difference is, however, two Uranian Trojans have been discovered, 2011 $QF_{99}$ and 2014 $YX_{49}$ (Alexandersen et al., 2014, de la Fuente & de la Fuente, 2017). Both of these large, up to 100 km plus objects are temporary captures and will move onto Centar like orbits within the next one to three million years.

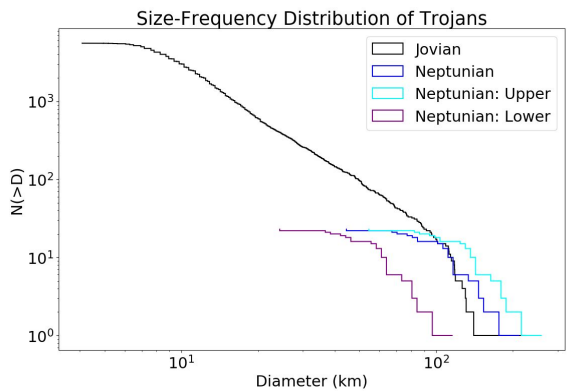

In contrast to the Saturnian and Uranian Trojan population, dynamical simulations have indicated that there should be a large, stable population of particular Neptunian Trojans (Nesvorny & Dones, 2002, Lykawka et al, 2010, Parker, 2015). So far, only 22 Neptunian Trojans have been discovered (Rui-jun et al., 2019). Though small in number, the orbits are indicative of a much larger population, possibly rivaling that of the Jovian Trojans. As with the Jovian population, several of the Neptunian Trojans show highly inclined (>25º) orbits, a 'warm', puffed up population. What this indicates is that the population was captured, rather than growing in place. This provides another, as yet understudied population of objects from the early Solar system. By contrasting these objects with the Jovian Trojans, the target of the *LUCY* spacecraft in the late 2020's (Levison et al. 2017), we can come to understand some of these capture mechanisms that occurred during the early Solar system. Also, since they are in the outer solar system, the Neptunian Trojans would have undergone less weathering compared with the Jovian Trojans, giving insight into these long term processes.

**Irregular Satellites of Uranus and Neptune**

Around each of the gas giants in the Solar system, there is a collection of small satellites that orbit at high inclinations. In the Jovian and Saturnian systems, these make up the majority of the number of satellites (e.g., Nicholson et al. 2008; Denk et al. 2018).

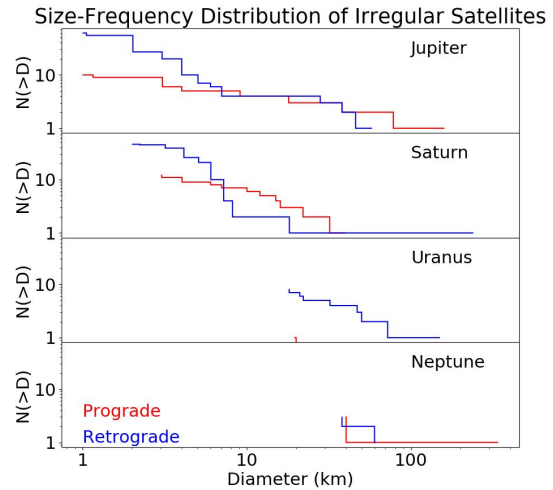

Each population can be split into two groups, the prograde satellites that orbit in the same direction as the planet and major satellites, and the retrograde objects that orbit in the opposite direction. Dynamically, these retrograde orbits are impossible to recreate with *in-situ* formation, like the Galilean satellites. It has been suggested, that like the Trojans the Irregular satellites were captured during a period of instability in the early Solar system (Nesvorny et al. 2007).

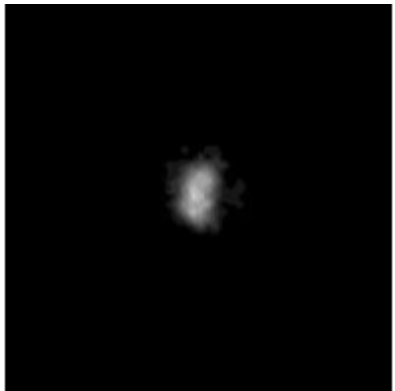

Current best image of an outer SS Irregular Satellite. Nereid, an ~357 km Neptunian from VOYAGER (NASA/JPL, 1996)

After capture, both the Jovian and Saturnian Irregular satellites went through a sequence of collisions (Bottke et al. 2010), that resulted in sets of families that can be seen today (Sheppard & Jewitt. 2003, Holt et al. 2018). In the Jovian and Saturnian systems, each family can be represented by the largest fragment. In the Uranian and Neptunian system, several of these larger Irregular satellites have been found (Jewitt & Haghighipour, 2007). In contrast to the Jovian and Saturnian systems, smaller members have not yet been discovered, likely because they are simply too faint (Sheppard et al. 2005, 2006). Uranus has nine Irregular moons, of which only one is prograde. All of these are larger than about 20 km, the size of the largest family fragments at Jupiter and Saturn. Neptune also has a small number of large Irregulars, three prograde and three retrograde (Holman et al. 2004), though more are expected to be found (Sheppard et al. 2006). The Neptunian system also houses Triton, a captured Edgeworth-Kuiper Belt object (Agnor & Hamilton. 2006) and itself of major interest, see Grundy et. al (2020) White Paper for more details.

**Links to early Solar system formation**

The captured small bodies, Trojans and Irregular satellites, provide vital clues as to the formation of our Solar system. They have unique dynamical situations, which any model of Solar system formation needs to explain. To date, the focus has solely been on the Jovian and Saturnian systems. The importance to understanding the early Solar system is one of the driving motivations behind the *LUCY* mission (Levison et al. 2017), to visit the Jovian Trojans in the late

2020s/early 2030s. One of the current canonical histories, the 'Nice' model, where Jupiter migration scatters the primordial disk (Nesvorny, 2018), was developed to explain the peculiar dynamics of the Jovian Trojans and Irregular satellites. By studying these objects in the context of the early Solar system, we can make links to Exoplanet systems (Horner et al. 2020).

## Current Unknowns and science questions

**What is the size distribution of the captured Small body populations in the Ice giant region?** The vast majority of the captured Small body population are currently unknown. A more complete size-frequency diagram will help inform the collisional environment in the outer Solar system. Size can be calculated from Albedo and magnitudes, both gathered from the NSF LSST and follow up programs. This is one of the primary outcomes from studies on small Solar system bodies in the next decade.

**What are the dynamical situations of the captured Small body populations in the Ice Giant region?** The dynamical stability of the Jovian Trojans has helped inform the history of the Solar system (Nesvorny, 2018). The contemporary Jovian Trojans drive the *n*-body simulations of the past population. As the contemporary population of the outer Solar system Trojans is unknown, it is difficult to simulate the past population (Parker, 2015).
Amongst the Jovian Trojans there have been several collisional families identified (Nesvorny et al. 2017). Currently, an insufficient number of objects have been found to identify any collisional families in the Trojan swarms of Uranus and Neptune.
With regards to the Irregular satellites, collisional groups have also been discovered (Sheppard, & Jewitt. 2003, Holt et al. 2018). Each group is represented by a large, type object (Holt et al. 2018). In the outer Solar system Irregular satellites, only these large objects have been discovered to date. In particular at Uranus, with its unusual axial tilt, understanding the dynamics of these objects will help inform the origin of the tilt, and the sequence of events related to the capture. At Neptune, two of the objects, Psamathe and Neso, may be members of a family (Sheppard et al, 2006), but discovery of additional members would be needed to confirm this.

**What are the colors and physical properties of the captured Small populations in the Ice Giant region?** The Jovian Trojan population consists mostly of D-type asteroids (DeMeo & Carry, 2013), under the current Bus-DeMeo taxonomy (DeMeo et al. 2009). The colours and spectra of the Jovain Trojans link them to the outer Kuiper Belt objects (Wong & Brown. 2016). Similar colours also link the Irregular satellites to the outer Solar system. These colours are currently poorly understood for captured small bodies in the Ice Giant region. Characterizing spectra and colours here will help to link these objects to other populations and help discover their origin.

**How do the captured small body populations in the Ice Giant region compare with equivalent populations?** These objects did not form in their current locations. Using dynamics and surveys, such as LSST, to characterize these objects, we can link them to other populations within the Solar system, such as Near Earth Objects, Centaurs, the Main and Edgeworth-Kuiper belts, as well as the Jovian Trojans and other Irregular satellites populations at Jupiter and Saturn. Followup from ground based institutes can place these objects in the context of the wider

Solar system. Discovering these links help form the theories of formation and evolution of small Solar system bodies.

**What can the captured small body populations in the Ice Giant region tell us about the history of the Solar system?** Jovian Trojans and Irregular satellites have informed our understanding of the early Solar system. Discovering the properties of these additional objects in the Ice Giant region can help put constraints on the current range of models.

## Recommendations

The captured objects in the Ice Giant region represent one of the last understudied populations in the Solar system. Previous studies on objects in the Jovian region have yielded a wealth of information about our Solar system history and driven a paradigm shift. The captured objects in the ice Giant region could similarly provide vital clues to our Solar system history. In the coming decade, NASA should support primary science objectives to discover and characterize these objects, in preparation for future surveys and missions.

**Provide support to discover additional captured objects in the Ice giant region:** There is a major gap in our knowledge of small captured objects in the Ice giant region, in particular the Neptunian Trojan population. It is recommended that additional time on large telescopes, such as BLAH, be allocated to discovering and categorizing these objects. The LSST will possibly discover several of these objects, though further modeling and analysis may be required to ascertain if the object is a permanent fixture of the populations, rather than from more transient populations, such as the Centaurs or Jupiter family comets. We support dedicated programs to fund analysis of LSST data under the NASA ROSES framework (LSST Science collaboration et al., 2020).

**Devote facility time to characterizing captured objects in the Ice giant region:** Once discovered, time will need to be devoted to characterizing these objects. Combined with expected low albedos, and distance, this gives approximate absolute magnitudes of 21.5 for the Uranian Trojans, and 23.3 to 23.7 for the Neptunians. These faint objects require access to the largest telescopes, which todate have only been used for discovery. These objects will be prime candidates for characterization using the next generation of ground and space-based telescopes, including JWST and the Thirty Meter Telescope (TMT), both expecting first light within the next decade.

**Incorporate captured outer Solar system bodies into mission narratives:** Discovery Phase A selected mission TRIDENT is currently exploring the prospect to observe Irregular Satellites of Neptune during its flyby through Neptune's system. Preliminary assessment indicates Neptune's second largest Irregular satellite Proteus (dia~420 km) may be observed from <70,000 km, which would be half the closest approach distance by Voyager 2. Combined with a better visible camera, this would lead to a spatial resolution of ~0.13 km/px or ~10x increase over the Voyager 2 observations (Howett, SBAG 23, 2020). Nereid (dia~357 km) would also be visible to Trident from ~5M km, resulting in an increase in spatial resolution of ~10 km/px or ~x4 increase over the Voyager 2 observations (Howett, SBAG 23, 2020). If the currently planned Neptune arrival time of Trident would be shifted forward by a few weeks, the minimum distance to

Nereid would shrink below 2M km. The other currently known Irregular moons of Neptune might also be observed by Trident with a strategy similar to the observations of Saturn's Irregulars performed by the Cassini spacecraft (Denk & Mottola 2019).

**Assist with technological development to study small outer Solar system objects:** As further information about these objects become available, new technologies will require development to further study these objects. Using the SIMPLEx program, small low cost missions can be developed to study objects in the size range of the captured small outer Solar system objects.. These would require testing in the NEO population, before potential missions to the Trojans or Irregular satellites of Uranus and Neptune could be undertaken in subsequent decades.

**Create missions and concept studies to study captured outer Solar system objects:** While many existing facilities can be used to investigate these objects, in the following decadal survey, new facilities and missions will be required. During this decadal, some preliminary studies will be required in preparation for the 2033 decadal call. Several NASA Small Solar system body missions, including DART, LUCY and PSYCHE will reach their respective targets within the next decade. As these missions reach their targets, the heritage instruments can be adapted to a potential future mission to one or more captuned objects in the Ice giant region. Concept studies into this potential mission during this decadal survey will ensure a successful future mission. NASA should provide funding for these preliminary studies in this decadal survey, in order to prepare for a proposal in the following 2033 decadal survey.

**Associated Whitepapers**
Buratti et al. (2020). The Small Satellites Of The Solar System: Priorities For The Decadal Study
Cartwright et al. (2020) The science case for spacecraft exploration of the Uranian satellites.
Davidsson et al. (2020), What do small bodies tell us about the formation of the Solar System and the conditions in the early solar nebula?
Grundy, et al. (2020), Triton: Fascinating moon, likely ocean world, compelling destination
Hofstadter et al. (2020) Exploration of the Ice Giants, Uranus and Neptune.
The LSST Solar System Science Collaboration. et al. (2020), The Scientific Impact of the Vera C. Rubin Observatory's Legacy Survey ofSpace and Time for Solar System Science
Milazzo et al. (2020), DEIA 101: Why is diversity important?
Omurhan et al. (2020) Exploration of Dwarf Planets, KBOs and Centaurs.
Rivera-Valentín et al. (2020), Who is Missing in Planetary Science?: Demographics showing Black and Latinx scientists are the most underrepresented
Rathbun et al. (2020), Who is Missing in Planetary Science?: Recommendations to increase the number of Black and Latinx scientists
Strauss et al. (2020), Non-binary inclusion in planetary science